\documentstyle[11pt]{article}
\date{}
\begin{document}
\sloppy
\title{The vacuum energy excitations due to gravitational field
as a possible candidate of dark matter in galaxies}
\author{V. Majern\'{\i}k \\
Institute of Mathematics, Slovak
Academy of Sciences, \\SK-814 73 Bratislava, \v Stef\'anikova  47,
 Slovak Republic\\ and\\
Department of Theoretical Physics\\ Palack\'y
University\\T\v r. 17. listopadu 50\\772 07 Olomouc, Czech Republic}
\maketitle
\begin{abstract}
\noindent
In this Letter we
point out to the possibility that the cloud of the
vacuum energy excitations in
gravitation fields surrounding galaxies forms a component
of dark matter. These clouds of the vacuum energy excitations interact
gravitationally
with the baryonic matter of galaxies changing their dynamical and
kinematical properties.
In four model galaxies
we show that the dynamic changes due to the vacuum energy excitations
of these galaxies are comparable with data.
This shows that vacuum energy excitations
created in the gravitation field of galaxies
 may be considered as one of
the candidates of dark matter \footnote{E-mail:majerv@prfnw.upol.cz}
\end{abstract}

\section{Introduction}

The recent astronomical observations 
 \cite{pe} \cite{BO} \cite{PER} give increasing
support for the 'cosmical concordance' model in which the universe is
flat and consists of a mixture of a small part of
baryonic matter about one third non-relativistic dark matter (DM) and two
thirds of a smooth component, called dark energy (DE).
Although the accelerate expansion of the universe appears to gain
further ground, the substratum behind of the dark energy
remains as elusive as ever \cite{2}.

In the literature,
DE is theoretically modelled by many ways, e.g. as
(i) a very small cosmological constant (e.g.\cite{4}) (ii) quintessence
(e.g.\cite{5})
(iii) Chaplygin gas (e.g.\cite{6}) (iv) tachyon field (e.g.\cite{7}) (v)
interacting quintessence (e.g.\cite{8})
non-minimally coupled Q (e.g.\cite{10}), quaternionic field (e.g.\cite{IK}),
etc.
It is unknown which of the said models will finally
emerges as the successful
one.

Another fundamental problem being faced to cosmology
is that of the nature of DM, which is supposed to exist because of
dynamical astronomical measurement but we have not yet detected it.
Astronomers found that one third of the universe's mass is made up of unknown
matter that is invisible to telescopes but have gravitational
effects on the baryonic matter \cite{D}.
Lacking evidence of direct detection, the presence, nature, and quantity
of DM must be inferred from the kinetic and distribution
properties of baryons.
In the literature, many sophisticated candidates of
DM have been proposed
which can be divided up into two categories. Some authors proposed
that DM consists of some kind of matter substance, e.g.
of the massive particles (WIMPs)
which stems from extensions to the standard model of particle physics,
such as supersymmetry and extra-dimensional theory. Other assume
that DM represents
the relics of
primordial black holes (see, e.g.\cite{F}).
On the other side, several authors try explain the kinematic effects
assigned to DM by modifying of Newton's law of
gravitation (see,e.g. \cite{P}).

Instead of looking for further sophisticated explanations of the nature
of DM
 we will, in what follows,
attempt to show that the same mechanism of the creation of matter in
the cosmic quaternionic field \cite{X} can also be applied to the matter
creation
in the weak gravitational field surrounding galaxies.
Recently, the field energy density of the so-called cosmical quaternionic
field
has been interpreted as the vacuum energy density and set equal to the
cosmological constant \cite{MK}.
It has be shown that in this field, due the action of a force field on the
virtual particles,
an spontaneous creation of matter occurs either in the
form of real particles or
the vacuum energy excitations.

In this Letter we apply
 a similar mechanism for matter creation also to gravitational field.
 We shown that due to the weakness of the galactic
gravitational field only  clouds of
the vacuum energy excitations
are created which interact gravitationally with the baryonic matter of
galaxies
changing their dynamical and kinetical properties.
Hence, we attempt to interpret
these clouds as a kind of DM  occurring
in the gravitational field of galaxies.

The Letter is organized as follows. In Section 2 we
describe the mechanism of the creation  of the vacuum energy
excitations in the weak gravitational field of galaxies.
Here, we present some model galaxies with the clouds of the
vacuum excitations and show that their dynamical parameters
assume the plausible values.
In Sections
we conclude that the vacuum energy excitations due to gravitation field
of galaxies might fulfil the requirements asked from  DM.

\section{The creation of vacuum energy excitations
in the weak gravitational field}

Motivated by the desire to find a possible candidate of the dark mass
we are aspirated by the matter creation in the
cosmic quaternionic field \cite{X}.
In \cite{IK} we have described a new mechanism of the creation of real
particles from the virtual ones in the presence of the cosmic
quaternionic field.
It is well-known that in absence of a force field in
vacuum the virtual particles
are created and after a time interval $\Delta
t=h/(c^2m_v$) again annihilated without being observed.
The basic idea here is that the virtual particles during the
time interval $\Delta t$ behave so as being real particles and due to the
interaction
with a force field they can gain a certain amount of energy.
Laking the consistent theory of the quantum
vacuum we use for the estimation of the
energy gains of virtual particles, during their
lifetime, the following simple heuristic arguments.
We start with fundamental equations of classical and quantum mechanics
$$\Delta E=F\Delta x \quad \eqno(1)$$
and
$$\Delta p \Delta x\geq h,\quad \eqno(2)$$
where $\Delta p$ is the uncertainty in the momentum of a virtual particle,
$\Delta x$ is the uncertainty in its
displacement and $F$ is the force acting on it.
Combining Eqs. (1) and (2) and taking for $\Delta p$ the rest momentum
of a virtual particle $m_vc$, we get for the gained energy the formula
$$\Delta E\approx \frac{Fh}{m_vc}.\quad \eqno(3)$$

For different force fields  $\Delta E$ assumes different values. E.g.
if we insert into Eq.(3) the force exerting on the mass body
in the cosmic quaternionic
field described in \cite{IK}
then $\Delta E$, gained from the ambient field, during
lifetime of virtual particles, is
$$F\Delta x= \Delta E \approx \sqrt{G} \Phi(t){h\over c} \approx {h\over
(t+t_0)},$$
where $\Phi(t)$ is the intensity of the cosmic quaternionic field.
If this intensity is sufficiently large, e.i. if $t$ and $t_0$
are short enough then it can separate
the virtual particles creating the real ones.
This happens in the vicinity of the Bing Bang \cite{MK}

If one writes the Newton law of gravitation
in the symmetrical form
$$F= \frac{(q_g)^2}{r^2}=\frac{(\sqrt{G}m)^2}{r^2},$$
then the quantity $q_g=\sqrt{Gm}$ represents
 the gravitational charge of the mass
$m$ \cite{BR} \cite{JM}. Next we will use gravitational charges when
formulating the common gravitation
equations.
As well-known the force acting on the mass $m_v$ in gravitation field
is given as
$$F_{g}=\sqrt{G}m_vE_g,$$
where $E_g=\nabla \phi$ is the 'intensity' od gravitational field given
by the gravitational potential $\phi$ and $\sqrt{G}m_v$ is the
gravitational charge of a virtual
particle. Taking for $\Delta p$ the rest
momentum of $m_v$, i.e. putting $\Delta x=h/m_vc,$
and inserting  Eq.(2) into Eq.(1) we obtain for $\Delta E_g$ the following
expression
$$\Delta E_g=F_{g}\Delta x=\sqrt{G}m_vE_g\Delta x
=\frac{\sqrt{G}hE_g}{c}=\frac{\sqrt{G}hE_g}{c}.$$
The corresponding mass is
$$\Delta m_g=\frac{Gh}{c^3}\frac{E_g}{\sqrt{G}}= \frac{L_p^2}{\sqrt{G}}
E_g,\quad \eqno(4)$$
The intensity of the gravitation field of galaxies $E_g$ and in
consequence also the gained energy $\Delta E_g$ is relatively small
creating only cloud of the vacuum energy excitations
which interact gravitationally
with the baryonic matter. The cloud of these energy excitations
surrounding a galaxy behave so as
a sort of DM which we will call the excitation matter.

 Next, we suppose that the mass density of the excitation
matter is proportional to $\Delta m_g$
$$\rho_e =\kappa \Delta m_g=\frac{\rho_p L^2_p \Delta m_g}{\sqrt{G}},$$
where $L_p=(Gh/c^3)^{1/2}=4.10^{-35}$ is the Plank length.
Here, $\rho_p=N/l^3$ where $N$ is number of energy excitations in the
volume $l^3$ and $\kappa=\rho_p L^2_p/\sqrt{G}$.
Since the expression $\rho_e L_p^2$
has the dimension $[m^{-1}]$ we set next
$$\rho_p L_p^2=\frac{1}{a},$$
where $a$ is an unknown free constant with the dimension $[m]$.
Hence, the  density of the excitation matter becomes
$$\rho_g(\vec r)=\frac{\rho_pL_p^2}{\sqrt{G}}E_g=\frac{E_g}{a\sqrt{G}}.$$

The gravitation intensity $\vec E_g(\vec r)$ of the gravitational field is
linked with the mass density by the
well-known equation
$$\nabla \vec E_g(\vec r) =\sqrt{G} \rho(\vec r), \quad \eqno (5)$$
where $\rho(\vec r)$ is mass density as a function of $\vec r$.
Since the excitation DM
of a galaxy is an additional source of its gravitation field
 Eq.(5) becomes
$$\nabla \vec E_g(\vec r) =\sqrt{G}\left (
\rho_b(\vec r) +\frac{E_g(\vec r}{a\sqrt{G}}\right ),\quad \eqno
(6)$$
where $\rho_b(\vec r)$ and $E_g(\vec r)/(a\sqrt{G})$
is the density of the baryonic and excitation matter, respectively.
Eq.(6) is the basic equation for the calculation of the density of the
excitation
matter surrounding galaxies.
Note that to determine this density
we {\it only} need the
density of its baryonic matter of galaxies.

The total baryonic and excitation matter $M_b$ and $M_e$ of a galaxy
having the volume $V$ is given as
$$M_b=\int_{V} \rho_b(\vec r) d\vec r$$
and
$$M_e=\int_V \frac{E_g(\vec r)}{a\sqrt{G}} d \vec r,$$
respectively.

Next, we confine ourselves to the spherically symmetrical case,
i.e. we suppose that the mass density of galaxy is dependent only on $r$.
Then Eq. (6) become a simple linear differential equation
$$\frac{dE_g(r)}{dr} +\frac{2E_g(r)}{r}=\sqrt{G}\left
(\rho_b(r)+\frac{E_g(r)}{a\sqrt{G}}\right )\quad\eqno(7)$$
whose general
solution has the form
$$E_g(r)=\exp^{-\int{(2/r-1/a) dr}}\left [\int {\rho_b (r)
\exp^{\int{(2/r-1/a) dr}}
dr} +C \right ]. $$
Given the density of the baryonic matter and the constant $a$
both $M_g$ and $M_e$ can be
calculated by means of Eq.(7). For the radius of a typical
galaxy we take $R=10^{21}m$. Since $a$ is a free constant we set it,
for the sake of simplicity, equal to $R$, i.e. $a=R=10^{21}m$, and with
such
chosen $a$ we calculate the baryonic and excitation masses of galaxies.
 Later we
will attempt to justify this value.

Let us, as an example, considered a galaxy with the density of the
bradyonic matter $\rho'(r)$ given by the formula
$$\rho'_g(r)=A\left (\frac{2}{r}-\frac{1}{a}\right), \qquad r\leq 2a$$
whose radius and the constant $a$ are equal to $10^{21} m$ and its total
bradyonic mass is $10^{42} kg$.
The solution of Eq.(7) for $\rho'_g(r)$ is simple
$\rho_e=A$.
By integrating $\rho_g$ and $\rho_e$ over the galaxy volume we obtain
for its total baryonic and excitation masses
$$M_g=\int_0^{R=a} \left (\frac{2}{r}-\frac{1}{a}\right ) 4\pi r^2 dr=
\frac{A16a^2\pi}{3}$$
and
$$M_e=\int_0^{R=a} A r^2 4\pi dr= \frac{A32a^2\pi}{3},$$
respectively.
The ratio of the total baryonic mass of a galaxy to its total excitation
mass, denoted by the symbol
$\alpha$, is given as
$$\frac{M_e}{M_g}=\alpha.$$
For $\rho'(r)$ it becomes equal to 2.
The total mass of a galaxy is given as the sum of both mass $M_g$ and
$M_e$
$$M_t=M_g+M_e$$
In the following Table we present
the characteristic quantities of galaxies, i.e. $M_g$, $M_e$ and
$\alpha$, with
the different densities of bradyonic matter presented in the first row
the following Table
\begin{center}
\begin{tabular}{|c|c|c|c|c|c|c|c|}
\hline
& n=1 & n=2&n=3& n=4 \\
\hline
$\rho_b(r) $ &  $A(2/r-1/a)$ &  $A(1/r-1/a)$ &
$A(1/r^2-1/a^2)$& $A(2/r^2-1/a^2)$\\
$E_g(r)$ & A &
$A(a^2+ar+r^2)/r^2$&$A(a^2+2ar+r^2)/r^2$&$A(2a+r)/ar$\\
$M_g$ & $(16a^2A\pi)/3$ &$(4\pi a^2A)/6$&$(8aA4\pi)/3$& $(16/3)\sqrt{2}A\pi$ \\
$M_e$ & $(32a^2A\pi/3$ & $(11 a^2
A4\pi)/6$&$(28aA\pi)/3$& $8aA\pi+(8/3)\sqrt{2}
aA\pi$\\
$M_e/M_g$&2 &11&3,5&4.9\\
$M_g+M_e$ & 3 & 12&4.5&5.9\\
\hline
\end{tabular}\\
\end{center}
\noindent
\section{Consequences}
It is generally believed that
most of energy and matter in our universe is of unknown nature to us,
therefore
the to explain  the
nature of DM is one of the major fundamental challenges
of present astrophysics. In this Letter we have attempted to show that the
cloud of the vacuum energy excitations in gravitation field of galaxies might
represent a form of DE.\\
From what has been said so far the following points are worth of
mentioning\\
(i) There exist a plausible mechanism for the creation of the vacuum energy
excitations
in the gravitation fields surrounding galaxies.\\
(ii) The density of the excitation matter is proportional to the intensity of
the corresponding gravitation field.\\
(iii) The constant $a$ depending on the number of energy
excitations in a unit volume, although very important for determining of
the excitation matter, cannot be determined exactly because of
laking the consistent theory quantum vacuum.\\
(iv) When setting $a=10^{21}m$, the total excitation mass of a galaxy
with the density of the bradyonic matter
given in Table is alway larger then its total baryonic mass.\\
(v) If we take for the volume, where an excitation in vacuum occurs,
the radius of nucleon $r_n\approx 10^{-15}-10^{-16}m$ then the number of
excitations in $[m^3]$ becomes $10^{45}-10^{48}$ which is equal approximately
the value $a=1/\rho_p L_P^2\approx 10^{24}-10^{21}m$.\\
(vi) The excitation DM can be characterized as a cloud of the vacuum energy
excitations created in the gravitation field,
neutral, undetectable by elm interaction with
negligible cross-section with bradyonic matter.
Hence, the cloud of the vacuum energy excitations
might represent a possible candidate for DM.

The aim of this short Letter was only to outline the basic idea of a
possible new form of dark matter. Many important issues remained open and
may be eventually solved by future consistent theory of quantum gravity
and using a more realistic model of galaxies. The relatively good
fit for the amount of the excitation DM surrounding
galaxies may be acceptable also from the purely
phenomenological point of view.

\end{document}